\documentclass[runningheads,anonymous]{llncs}

\usepackage[T1]{fontenc}

\usepackage{amsmath}
\usepackage{algorithm}
\usepackage{algpseudocode}
\newcommand{\algname}[1] {{\fontfamily{cmtt}\selectfont {#1}}}
\usepackage{dsfont}

\usepackage{booktabs}
\usepackage{caption}
\usepackage{subcaption}
\usepackage{multirow}
\usepackage{graphicx,wrapfig,lipsum}

\usepackage{float}

\usepackage{bbm}
\usepackage{array}
\newcommand{\header}[1]{\vspace*{1mm}\noindent\textbf{#1}.}

\usepackage[absolute]{textpos}

\begin{document}

\begin{textblock}{10}(3,0.4)
 \noindent\normalsize  \begin{center}International Workshop on Recommender Systems for Sustainability and Social Good \\ in conjunction with ACM RecSys 2024\end{center}
 \end{textblock}

\title{Correcting Popularity Bias in Recommender Systems via Item Loss Equalization}

\titlerunning{Correcting for Popularity Bias in Recommender Systems via ILE}

\author{Juno Prent\inst{1} \and
Masoud Mansoury\inst{2}
}
\authorrunning{Juno Prent and
Masoud Mansoury}

\institute{University of Amsterdam, Amsterdam, Netherlands\\
\email{juno.prent@student.uva.nl}\\
\and
Delft University of Technology, Delft, Netherlands\\
\email{m.mansoury@tudelft.nl}
}
\maketitle

\begin{abstract}
    Recommender Systems (RS) often suffer from popularity bias, where a small set of popular items dominate the recommendation results due to their high interaction rates, leaving many less popular items overlooked. This phenomenon disproportionately benefits users with mainstream tastes while neglecting those with niche interests, leading to unfairness among users and exacerbating disparities in recommendation quality across different user groups. In this paper, we propose an in-processing approach to address this issue by intervening in the training process of recommendation models. Drawing inspiration from fair empirical risk minimization in machine learning, we augment the objective function of the recommendation model with an additional term aimed at minimizing the disparity in loss values across different item groups during the training process. Our approach is evaluated through extensive experiments on two real-world datasets and compared against state-of-the-art baselines. The results demonstrate the superior efficacy of our method in mitigating the unfairness of popularity bias while incurring only negligible loss in recommendation accuracy.

\end{abstract}



\section{Introduction}


Popularity bias is a well-known issue in recommender systems which refers to the fact that a few popular items are frequently interacted with by the user, while the majority of other less popular items are rarely interacted with~\cite{abdollahpouri2017controlling,ahanger2022popularity}. This bias results in a number of issues including unfair representation of items in the recommendation results~\cite{singh2018fairness,mansoury2022understanding,mansoury2024mitigating,liu2024measuring,mansoury2021fairness}, mainstream bias~\cite{zhu2022fighting,li2021leave,li2023mitigating,pan2024countering}, and disparate recommendation quality across users~\cite{li2021user,naghiaei2022unfairness,huang2024going}. All these issues happen because the recommendation algorithm propagates and intensifies the existing bias in the input data to the recommendation output, what is known as \textit{algorithmic bias}~\cite{mansoury2020feedback,edizel2020fairecsys}. 


During the training process of a recommender system, the focus is primarily on obtaining recommendations that globally reach the highest possible performance. This is done by minimizing the loss function for each batch of user-item pairs during the training phase. Although this will lead to the best performance across all users, it may lead to a disparity between different groups of users. As the model is incentivized to obtain the best performance, it might discover that it can best do so by for instance not bothering to try learning when to recommend niche items to niche users, as these occur only rarely. This makes it easier for the model to learn which items to predict for users with mainstream interests, which is often the largest segment of users, but leaves a lot to be desired for the more niche segments of items. This can lead to stark differences between the training losses for different "mainstreamness" groups of users and items.


Our analysis in this paper reveals that popularity of the items negatively correlates with the loss values that those items receive during the training process: popular items reach the lowest loss values, while less popular items require more iterations to reach the optimum. The possible explanation for this behavior is that there are many more interactions on the popular items than the less popular items, hence more training iterations and updates are performed on the embeddings corresponding to those items which results in faster convergence rate for those popular items.

To address this issue of disparate loss across different items, we propose a new constraint, \textit{item loss equalization (ILE)}, and integrate it into the objective function of the recommendation model. During the training process of the recommendation model, together with the original objective function that optimizes for the best overall performance, ILE minimizes the disparity of loss values across different item groups. This ensures a commensurate optimization process across different items which can be translated as obtaining the same quality embeddings for items.
Extensive experiments using Bayesian Probabilistic Ranking (BPR)~\cite{bpr} on three publicly-available datasets and comparison with baselines show that our method improves the fairness of the recommendations for users and items, while maintaining the ranking quality of the recommendations. 

\section{Related work}

Fairness has recently been the topic of extensive research in recommender systems~\cite{wang2023survey,chen2023bias,mansoury2020feedback}. The principles of fair recommendations are often compromised due to inherent biases either in the input data or within the recommendation model itself. One of the most prevalent biases in this context is popularity bias~\cite{abdollahpouri2017controlling,zhu2021popularity}.

Most research addressing popularity bias focuses on either in-processing approaches where the training process or underlying objective function of the recommendation model is modified for fair recommendation generation, or post-processing approaches where long recommendation lists are produced and processed for fairer recommendation outcomes. Huang et al.~\cite{huang2024going} proposed a solution for multifactorial bias, a combination of popularity and positivity bias, by formulating an appropriate propensity function and incorporating it into the objective function of the recommendation models. Abdollahpouri et al.~\cite{abdollahpouri2017controlling} addressed popularity bias through incorporating a fair regularization term into the objective function of the recommendation model to balance the representation of different items (in particular popular versus unpopular items) in the recommendation lists.

On the other hand, in literature on post-processing approaches, Abdollahpouri et al.~\cite{calibrated_popularity} tackled this issue by calibrating the recommendation results with respect to the users' interest toward popularity spectrum. Antikacioglu and Ravi in~\cite{antikacioglu2017post} reduced the effect of popularity bias by improving the exposure fairness for items. This is done by solving a minimum cost flow problem on the bipartite recommendation graph. Singh and Joachims in~\cite{singh2018fairness} proposed a post-processing approach that achieves the fairness goals by solving a constrained objective function using linear programming. Zhu et al~\cite{zhu2021popularity} proposed solutions in both directions, in-processing and post-processing. In their in-processing approach, similar to~\cite{abdollahpouri2017controlling}, a regularization-based approach was proposed to mitigate the effect of popularity bias, while in their post-processing approach, a popularity compensation solution was employed to adjust the predicted relevance score according to the popularity spectrum of the items and the users' preferences, e.g., promoting less popular items by compensating for them more. Mansoury et al. in~\cite{mansoury2023fairness,mansoury2021unbiased} studied popularity bias in dynamic environment where the recommendation system is operating over time and in~\cite{mansoury2023potential} studied the factors leading to popularity bias in recommender systems.

A relevant line of research involves addressing mainstream bias, which is closely linked to popularity bias. Mainstream bias refers to the fact that users with interest toward mainstream items (often popular ones) are dominantly represented by the recommendation models and their preference negatively affects the recommendations to the non-mainstream users~\cite{pan2024countering}. In tackling this bias, Zhu and Caverlee~\cite{zhu2022fighting} addressed this bias through local fine-tuning, while Li et al.~\cite{li2023mitigating} employed cost-sensitive learning to mitigate the effects of mainstream bias.

\section{THE PROPOSED APPROACH}

We denote the set of users as $U=\{u_1,...,u_n\}$ and the set of items as $I=\{i_1,...,i_m\}$. In implicit feedback scenarios, interactions between users and items are recorded in a binary matrix $R \in \{0,1\}^{n \times m}$ where each entry $R_{ui} =1$ if user $u$ has interacted with item $i$ and $0$ otherwise. Following the item groupings in \cite{calibrated_popularity}, we group the items into three different segments based on their popularity degree: \textbf{Head items (H)} are the 20\% of the most popular items, \textbf{Tail items (T)} are the 20\% of the least popular items, and \textbf{Mid items (M)} are the rest of the items. 

\begin{figure}
    \centering
    \begin{subfigure}{0.33\linewidth}
      \includegraphics[width=\linewidth]{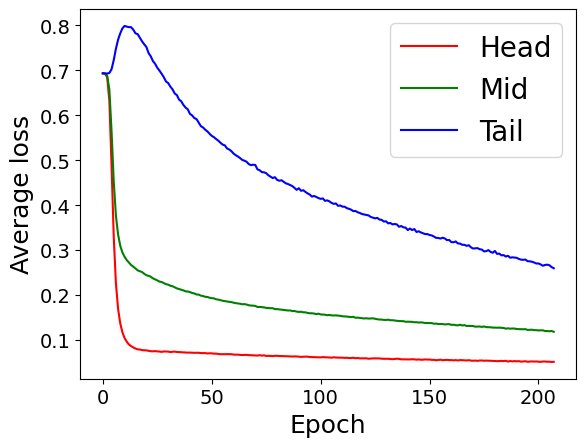}
      \caption{MovieLens1M}\label{fig: item group losses ml}
    \end{subfigure}\hfill
    \begin{subfigure}{0.33\linewidth}
      \includegraphics[width=\linewidth]{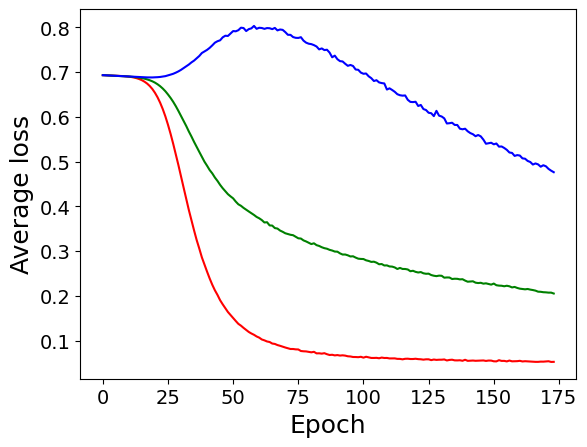}
      \caption{Goodreads}\label{fig: item group losses gr}
    \end{subfigure}\hfill
    \begin{subfigure}{0.33\linewidth}
      \includegraphics[width=\linewidth]{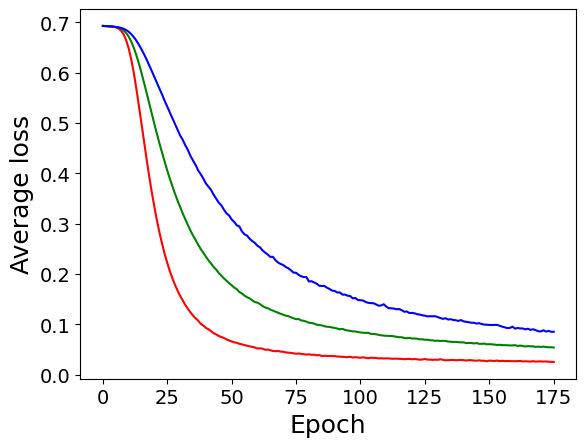}
      \caption{Google Reviews}\label{fig: item group losses go}
    \end{subfigure}\hfill
    
    \caption{Training losses per item group.}
    \label{fig: item group losses}
\end{figure}

\subsection{Ranking optimization}

In a ranking recommendation task, the goal is to predict whether a target user is interested in interacting with an item or not. This is done by minimizing a loss function that compares positive samples $s_{ui}$ (interacted items) with negative samples $s_{uj}$ (unseen items):
\begin{equation}\label{eq_ranking_obj}
    \mathcal{L} = \sum_{(u,i,j) \in D}{\delta(s_{ui},s_{uj})}
\end{equation}
\noindent where $\delta$ is a comparison function that returns the probability score of item $i$ being more preferable than item $j$ for user $u$ and can be in various forms, such as a sigmoid function.

This loss function iterates over the user-item pairs in interaction data $R$ and computes the loss value at each iteration according to the predicted probability score. Then, the model parameters are updated based on the computed loss value for the corresponding user-item pairs. This process continues until the model converges, i.e., the loss value across all user-item pairs no longer changes significantly.

The problem with this training process is that the process of computing the loss value and updating the parameters is not commensurate across different items. Due to the popularity bias in the interaction data, the popular items are more frequently selected in Eq. \ref{eq_ranking_obj} as much more interactions are observed on them. As a result, the model parameters corresponding to those popular items will faster converge than those less popular items. Figure \ref{fig: item group losses} shows the loss values across different item groups during the training of Bayesian Probabilistic Ranking (BPR) algorithm~\cite{bpr} on MovieLens, Goodreads, and Google Reviews datasets. As shown, although the model globally seems to converge after 200 epochs on MovieLens1M and 175 epochs on Goodreads and Google Reviews, there is a high disparity in the convergence and loss values for different items groups. Also, it is evident that while H items are quickly converged after a few iterations, T items still require many more iterations to optimize. Therefore, a treatment is needed to balance the loss across different item groups during the training process. 

\begin{table}[t]
\caption{Various distance functions ($\mathcal{D}$) in Eq. \ref{eq_loss_ile}. For all these functions, lower value signifies fairer output. The notation $|x|$ returns the absolute value of scalar $x$ and $||X||$ returns the size of vector $X$.}
\label{tbl_dist_funcs}
\centering
\begin{tabular}{@{}lll@{}}
\toprule
\textbf{Function}                & \textbf{Abbreviation} & \textbf{Formula}
\\ 
\toprule
Standard deviation     & STD & $\sqrt{\frac{\sum_{g \in G}{(L_g-\overline{L})^2}}{||G||}}$      
\\
\midrule
Entropy & ENT     &  $-\sum_{g \in G}{L_g . \log{L_g}}$      
\\
\midrule
Mean average deviation     & MAD & $\frac{\sum_{g \in G}{|L_g-\overline{L}|}}{||G||}$       
\\ 
\bottomrule
\end{tabular}

\end{table}

\subsection{Item Loss Equalization}
To address the issue of disparate loss values across different item groups during the training process, inspired by the research works in~\cite{williamson2019fairness,oneto2020general} in the area of fair machine learning, we introduce a constraint and integrate it into the objective function of the recommendation model. The goal of this new constraint is to minimize the disparity of aggregated loss values among different item groups. Hence, we rewrite Eq. \ref{eq_ranking_obj} as follows:  
\begin{equation}\label{eq_loss_ile}
    \mathcal{L}^*= \mathcal{L} + \lambda \times \mathcal{D}(\{L_g|g \in G\})
\end{equation}

\noindent where $\mathcal{L}$ is the average loss over all the training interactions as in Eq. \ref{eq_ranking_obj}, $\mathcal{D}$ is a distance function, $L_g$ is the average loss of items in group $g$, $G$ are the item groups, here defined as $G=\{H,M,T\}$, and $\lambda$ is a hyperparameter that controls the degree of focus on the original loss function or the newly added constraint. The higher the value of $\lambda$, the more the emphasis shifts to enforcing fair optimization. We call the new loss function in Eq. \ref{eq_loss_ile} \textit{Item Loss Equalization (ILE)}.

This new constraint forces the optimization process to strive for not only global performance, but also the minimization of the training loss differences between the groups of items. This is intended to have the model make predictions of similar quality for different groups of items. ILE is general and can be applied to any model-based recommendation system.

A core part of ILE is the distance function that computes how close the loss values corresponding to different item groups are. The more similar the groups' average losses are, the lower the distance would be. This is also consistent with the goal of the model, as it tries to minimize its overall loss values. As the groups' item losses getting closer during the training process, the model's predictions are more similar for different item groups, making the recommendation process more fair. Table \ref{tbl_dist_funcs} summarizes various distance functions that we experimented with in this paper.

\section{Experiments}
In this section, we describe the datasets, baselines, metrics, and experimental setting used for our experiments.
\subsection{Datasets}
We perform our experiments on three publicly-available datasets:

\begin{itemize}
    \item \textbf{MovieLens1M}\footnote{\url{https://grouplens.org/datasets/movielens/1m/}} is a dataset describing historical interactions by people (users) with movies (items). It contains 1,000,209 interactions, split across 6,040 users and 3,706 items. When used with the base BPR model, it shows an inherent training bias towards popular items, as seen in Figure \ref{fig: item group losses ml}.
    \item \textbf{Goodreads}\footnote{\url{https://github.com/BahramJannesar/GoodreadsBookDataset}} is a dataset describing historical interactions by people (users) with books (items). It contains 137,045 interactions, split across 2,225 users and 3,423 items. When used with the base BPR model, it shows an inherent training bias towards popular items, as seen in Figure \ref{fig: item group losses gr}.
    \item \textbf{Google Reviews}\footnote{\url{https://jiachengli1995.github.io/google/index.htm}} is a dataset describing historical interactions by people (users) with businesses (items). It contains 391,715 interactions, split across 12,791 users and 16,101 items. When used with the base BPR model, it does not show a strong inherent training bias towards popular items, as seen in Figure \ref{fig: item group losses go}.
\end{itemize}


\subsection{Baselines}
\label{section:baselines}
We compare our ILE method with the following baselines:

\header{Calibrated Popularity (CP)~\cite{calibrated_popularity}} This is a post-processing approach that processes the long recommendation lists by greedily minimizing the distance between the distribution of the $(H,M,T)$ items in the user's profile in input data with this distribution in recommendation lists. This way, it ensures that users interested in less popular items will receive less popular items in their recommendation lists. The size of the long recommendation lists ($N$) affects the performance of CP: shorter lists will result in lower fairness and lower accuracy loss, while larger lists lead to higher fairness and higher accuracy loss. Another hyperparameter in CP is $\lambda$ which controls the degree of emphasis on fairness improvement.

\header{Predictive Uncertainty-based Fair Ranking (PUFR) \cite{uncertainty}} This baseline utilizes the uncertainty of the recommendation model in predicting the relevance scores to improve the fairness. This is done by increasing the predicted relevance scores of the protected items and reducing the predicted relevance score of the unprotected scores by their uncertainty value. This way, PUFR ensures that the re-ranking would only happen within the range of model certainty to avoid significant loss in accuracy. 

\header{Inverse Propensity Scoring (IPS) \cite{schnabel2016recommendations}} This
approach directly alters each user-item pair’s loss value during the training process based on the
related item’s popularity, which is why it is an in-processing bias mitigation method. It does
so to atone for any item popularity bias within the used dataset. As IPS makes adjustments
to a model’s existing loss function, it is closely related to the principle behind ILE.
    
In our experiment, we approximate the model uncertainty by computing the standard deviation over the predicted relevance scores for each item when the model is trained with varying initialization seeds. We first train the recommendation model with 5 different randomly selected seeds. Each model predicts a relevance score for each user-item pair. The average score per item for each run is taken, and the standard deviation of these five average scores are computed as the uncertainty of model for the corresponding item. We also define $T$ items as the protected group and $H$ items as the unprotected group. Similar to CP, PUFR also involves a hyperparameter $\lambda$ that controls the degree of fairness improvement.

\subsection{Evaluation metrics}

We evaluate the ranking quality of our ILE method and baselines by Normalized Discounted Cumulative Gain (nDCG). We also use the following metrics for measuring the fairness of each method:

\header{User Popularity Deviation (UPD) \cite{calibrated_popularity}} This metric measures how well the distribution of the items groups in a user's profile match this distribution in user's recommendation list. Given a user's profile with the interacted items, we form the popularity distribution with the ratio of $(H,M,T)$ items in her profile. Similarly, we create the popularity distribution on her recommendation list. Then, we compute Jensen-Shannon Divergence (JSD)~\cite{js-div} between the two distributions. UPD measures the average JSD across all users. It outputs a value between 0 and 1, where 0 indicates fairest results.

\header{Aggregate Diversity (AD) \cite{antikacioglu2017post}} This metric measures the percentage of items that appear at least once in the recommendation lists.
    
\header{Equality of Exposure (EE) \cite{raj2022measuring}} Since an item can appear in the recommendation lists of different users and for each user can also be shown at different positions in the list, this metric measures how equal all items are represented in the recommendation lists. Following \cite{singh2018fairness}, we compute the exposure of an item $i$ according to its position in the list as $\frac{1}{1+\log{k}}$ where $k$ is the position of $i$ in the list. This forms a distribution of exposure across all items where uniform distribution signifies fairest results. To measure the uniformity of exposure distribution, we compute the Gini Index over this distribution. Given that the Gini Index falls within the range of $[0,1]$, for consistency, we report 1 minus the Gini Index for this metric. Consequently, EE of 1 signifies the fairest outcome (indicating a perfectly uniform distribution), while EE of 0 denotes the most unfair outcome.


\subsection{Experimental setup}

We randomly divide users' profiles into training and test sets, with 80\% allocated to the training set and 20\% to the test set. The training set is used to train the recommendation model and the test set is used to evaluate the recommendation lists.

While our ILE method is general and can be applied to any model-based recommendation algorithm, we base our experiments in this paper on Bayesian Probabilistic Ranking (BPR) algorithm~\cite{bpr}. We tune BPR by performing grid-search on its hyperparameters and report the best-performing results which is obtained for learning rate $10^{-4}$ and embedding size 128. The number of epochs is based on the convergence of the algorithm.
We generate a recommendation list of size $10$ for each user. We used RecBole\footnote{\url{https://github.com/RUCAIBox/RecBole}} for running experiments. 

\begin{table*}[t]
    \centering
    \caption{Performance comparison of our \algname{ILE} method with baselines in terms of ranking quality and fairness metrics. The bolded and italicized entries show the best and second-best values in each metric.} %
    \aboverulesep=0ex
    \setlength{\tabcolsep}{9pt}
    \belowrulesep=0ex
    \label{tab:results}%
    \centering
    \begin{tabular}{l| l| l l l l l}
        \toprule
        Method & Parameters & nDCG$\uparrow$ & UPD$\downarrow$ & AD$\uparrow$ & EE$\uparrow$ \\
        \midrule
        \multicolumn{6}{c}{\textbf{MovieLens1M}} \\
        \midrule
        \algname{BPR} & - & \textbf{0.2686} & 0.1208 & 0.4228 & 0.0963 \\
        \algname{CP} & $\lambda=0.98$, $N=50$ & 0.2263 & 0.0987 & 0.4096 & 0.0839 \\
        \algname{CP} & $\lambda=0.98$, $N=100$ & 0.2219 & 0.0925 & 0.4077 & 0.0827 \\
        \algname{CP} & $\lambda=0.98$, $N=200$ & 0.2213 & 0.0913 & 0.4077 & 0.0825 \\
        \algname{PUFR} & $\lambda=4.0$ & 0.2341 & 0.0853 & 0.4436 & 0.0908 \\
        \algname{IPS} & - & 0.2358 & 0.0853 & 0.4838 & \textit{0.1462} \\
        \midrule
        \algname{ILE} & $\lambda=0.25$, $\mathcal{D}=STD$ & 0.2263 & \textbf{0.0674} & \textit{0.4884} & 0.13 \\
        \algname{ILE} & $\lambda=0.04$, $\mathcal{D}=ENT$ & 0.2239 & \textit{0.0697} & \textbf{0.5245} & \textbf{0.1509} \\
        \algname{ILE} & $\lambda=0.3$, $\mathcal{D}=MAD$ & \textit{0.2413} & 0.0931 & 0.4444 & 0.1031 \\

        \midrule
        \multicolumn{6}{c}{\textbf{Goodreads}} \\
        \midrule
        \algname{BPR} & - & \textbf{0.133} & 0.2966 & 0.1341 & 0.0177 \\
        \algname{CP} & $\lambda=0.97$, $N=50$ & 0.112 & 0.2835 & 0.1394 & 0.0174 \\
        \algname{CP} & $\lambda=0.98$, $N=100$ & 0.1082 & 0.2727 & 0.1414 & 0.0176\\
        \algname{CP} & $\lambda=0.97$, $N=200$ & 0.1107 & 0.2784 & 0.1402 & 0.0176\\
        \algname{PUFR} & $\lambda=6.0$ & \textit{0.129} & 0.2212 & 0.1405 & 0.0248 \\
        \algname{IPS} & - & 0.1244 & 0.188 & \textbf{0.1805} & \textbf{0.0367} \\
        \midrule
        \algname{ILE} & $\lambda=0.3$, $\mathcal{D}=STD$ & 0.1034 & \textbf{0.1036} & 0.1543 & 0.0258 \\
        \algname{ILE} & $\lambda=0.02$, $\mathcal{D}=ENT$ & 0.1208 & \textit{0.1371} & 0.1107 & 0.013 \\
        \algname{ILE} & $\lambda=0.4$, $\mathcal{D}=MAD$ & 0.1249 & 0.147 & \textit{0.1633} & \textit{0.0293} \\

        \midrule
        \multicolumn{6}{c}{\textbf{Google Reviews}} \\
        \midrule
        \algname{BPR} & - & 0.0592 & 0.3255 & 0.1351 & 0.0253 \\
        \algname{CP} & $\lambda=0.99$, $N=50$ & 0.0536 & 0.3112 & 0.1458 & 0.0272 \\
        \algname{CP} & $\lambda=0.98$, $N=100$ & 0.0536 & 0.3101 & 0.1448 & 0.0272 \\
        \algname{CP} & $\lambda=0.98$, $N=200$ & 0.0536 & 0.3096 & 0.1452 & 0.0273 \\
        \algname{PUFR} & $\lambda=6.0$ & 0.0596 & 0.2866 & 0.1553 & 0.0291 \\
        \algname{IPS} & - & 0.0636 & 0.2919 & 0.1756 & 0.0367 \\
        \midrule
        \algname{ILE} & $\lambda=0.25$, $\mathcal{D}=STD$ & \textbf{0.0673} & 0.1996 & \textit{0.2886} & \textit{0.0625} \\
        \algname{ILE} & $\lambda=0.03$, $\mathcal{D}=ENT$ & 0.0586 & \textbf{0.1799} & 0.1905 & 0.038 \\
        \algname{ILE} & $\lambda=0.5$, $\mathcal{D}=MAD$ & \textit{0.0665} & \textit{0.1949} & \textbf{0.2932} & \textbf{0.064} \\
        
        \bottomrule
   \end{tabular}
    \vspace{-10pt}
\end{table*}%

\begin{figure*}[t]
  \centering
    \begin{subfigure}{\linewidth}
      \includegraphics[width=\linewidth]{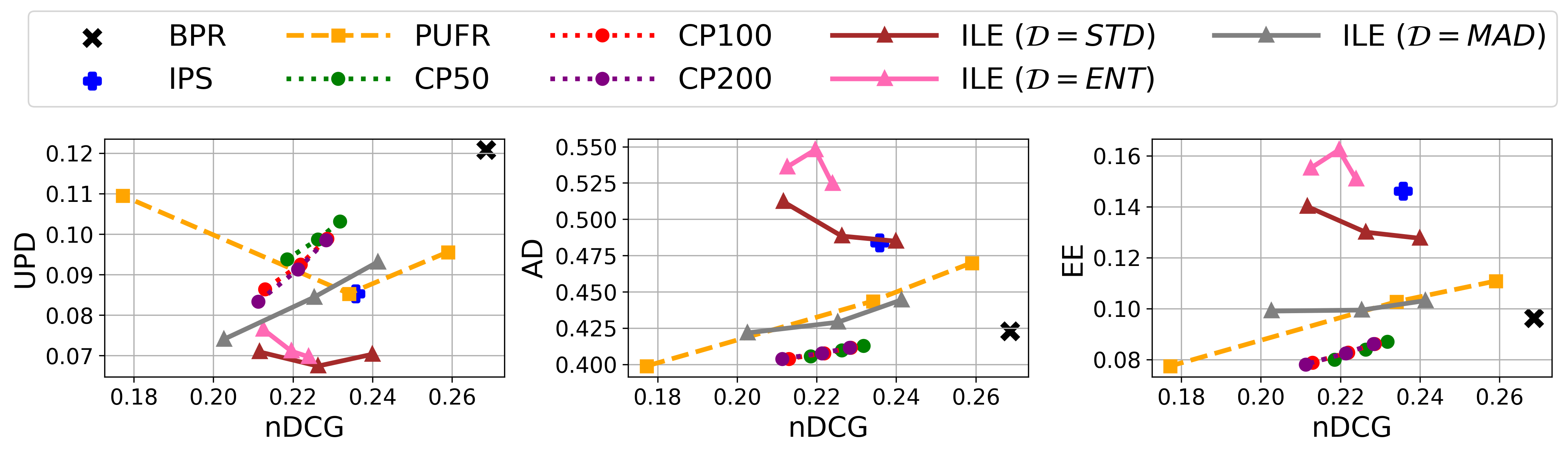}
      \caption{MovieLens1M}\label{fig:fairness-performance ml1m}
    \end{subfigure}
    \begin{subfigure}{\linewidth}
      \includegraphics[width=\linewidth]{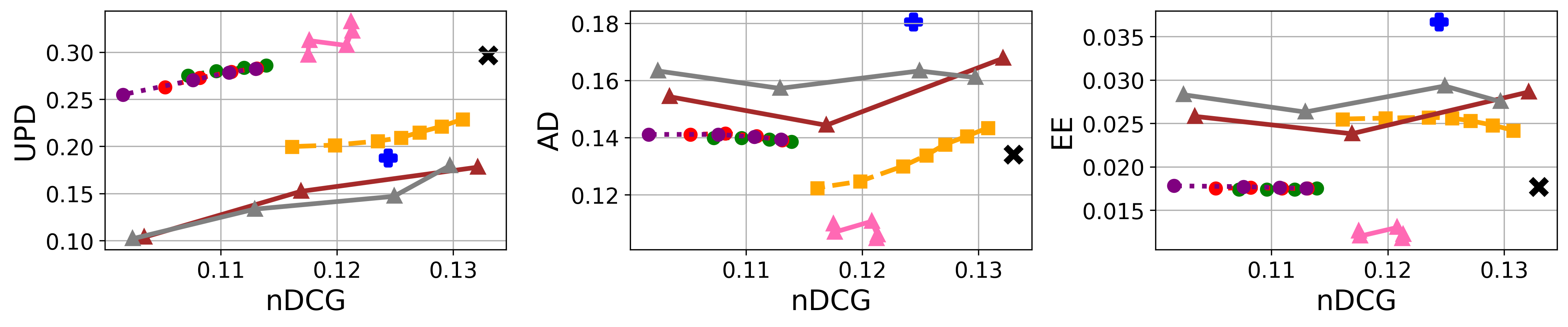}
      \caption{Goodreads}\label{fig:fairness-performance goodreads}
    \end{subfigure}
    \begin{subfigure}{\linewidth}
      \includegraphics[width=\linewidth]{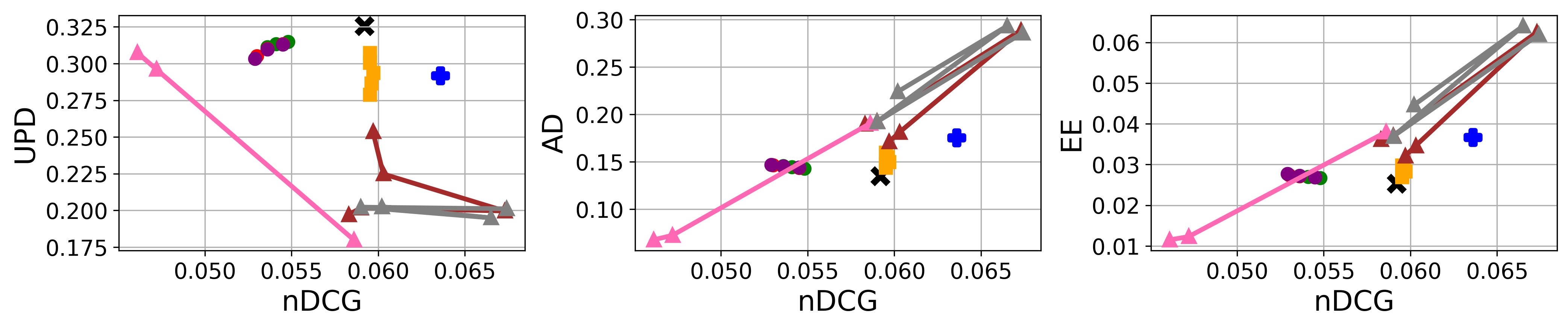}
      \caption{Google Reviews}\label{fig:fairness-performance googlereview}
    \end{subfigure}
    \caption{Trade-off between nDCG and fairness metrics for varying $\lambda$ values on each method.}
    \label{fig:fairness-performance}
\end{figure*}




\begin{figure}
  \centering
    \begin{subfigure}{0.33\linewidth}
      \includegraphics[width=\linewidth]{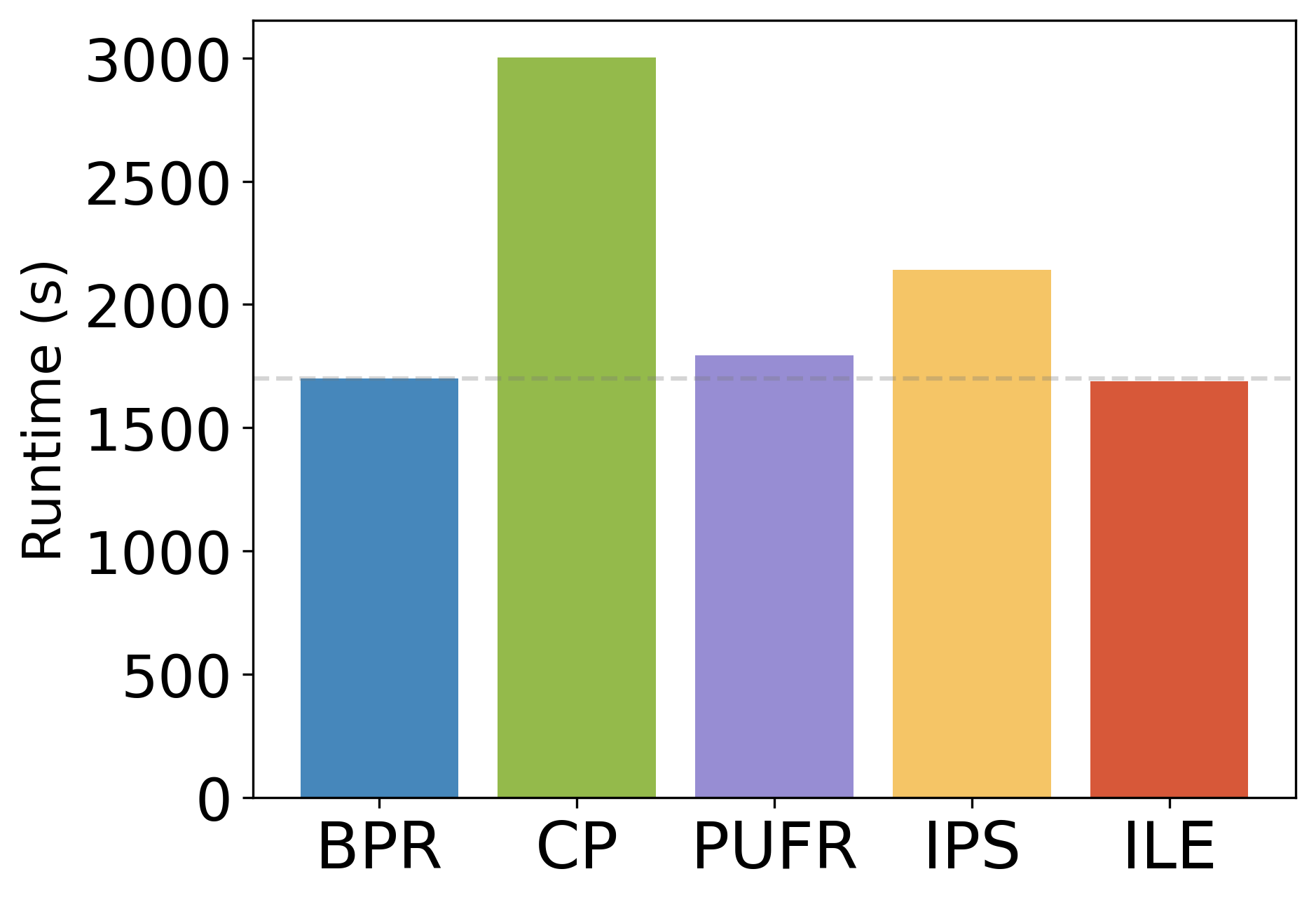}
      \caption{MovieLens1M}\label{fig:runtimes ml1m}
    \end{subfigure}\hfill
    \begin{subfigure}{0.33\linewidth}
      \includegraphics[width=\linewidth]{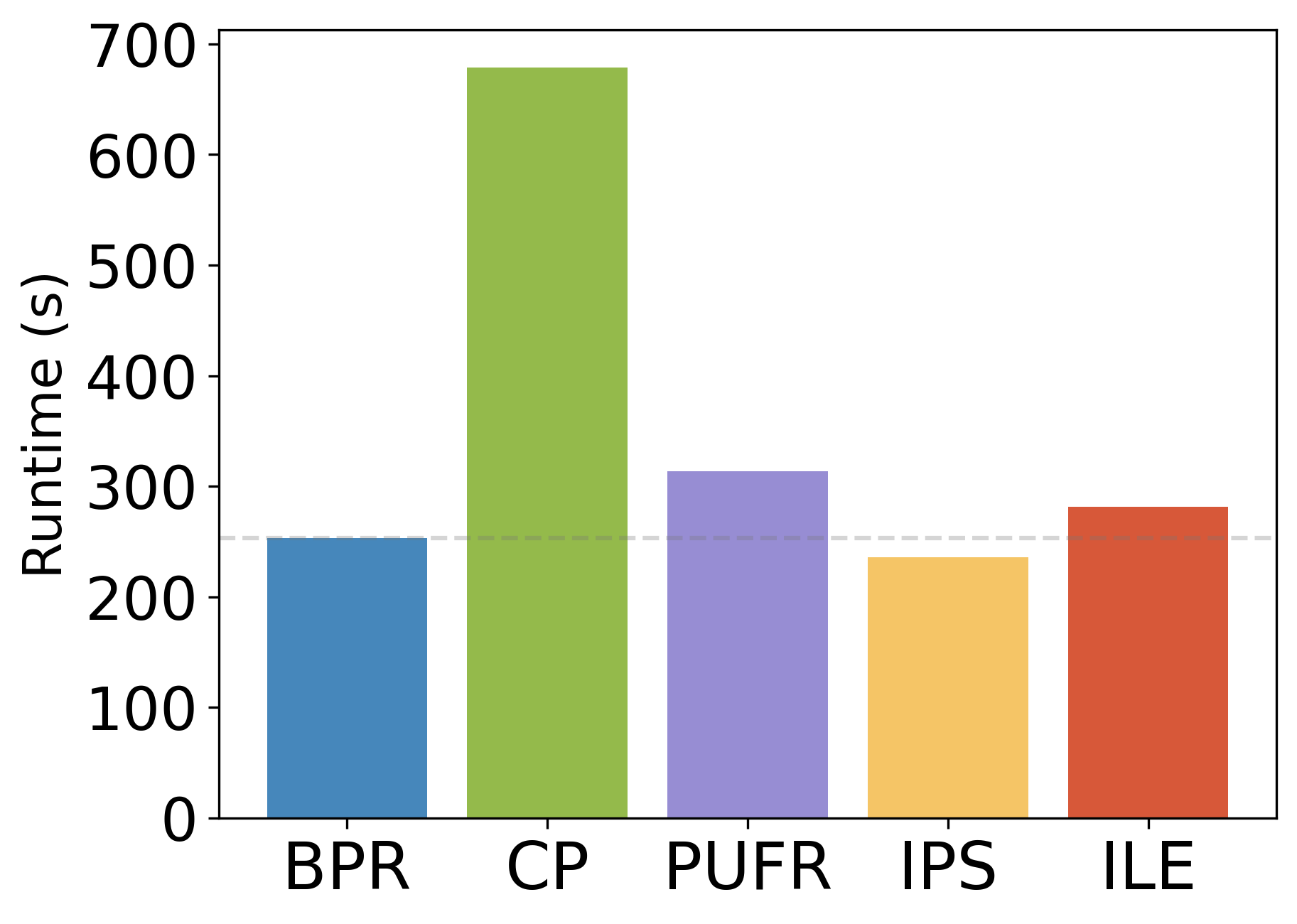}
      \caption{Goodreads}\label{fig:runtimes goodreads}
    \end{subfigure}\hfill
    \begin{subfigure}{0.33\linewidth}
      \includegraphics[width=\linewidth]{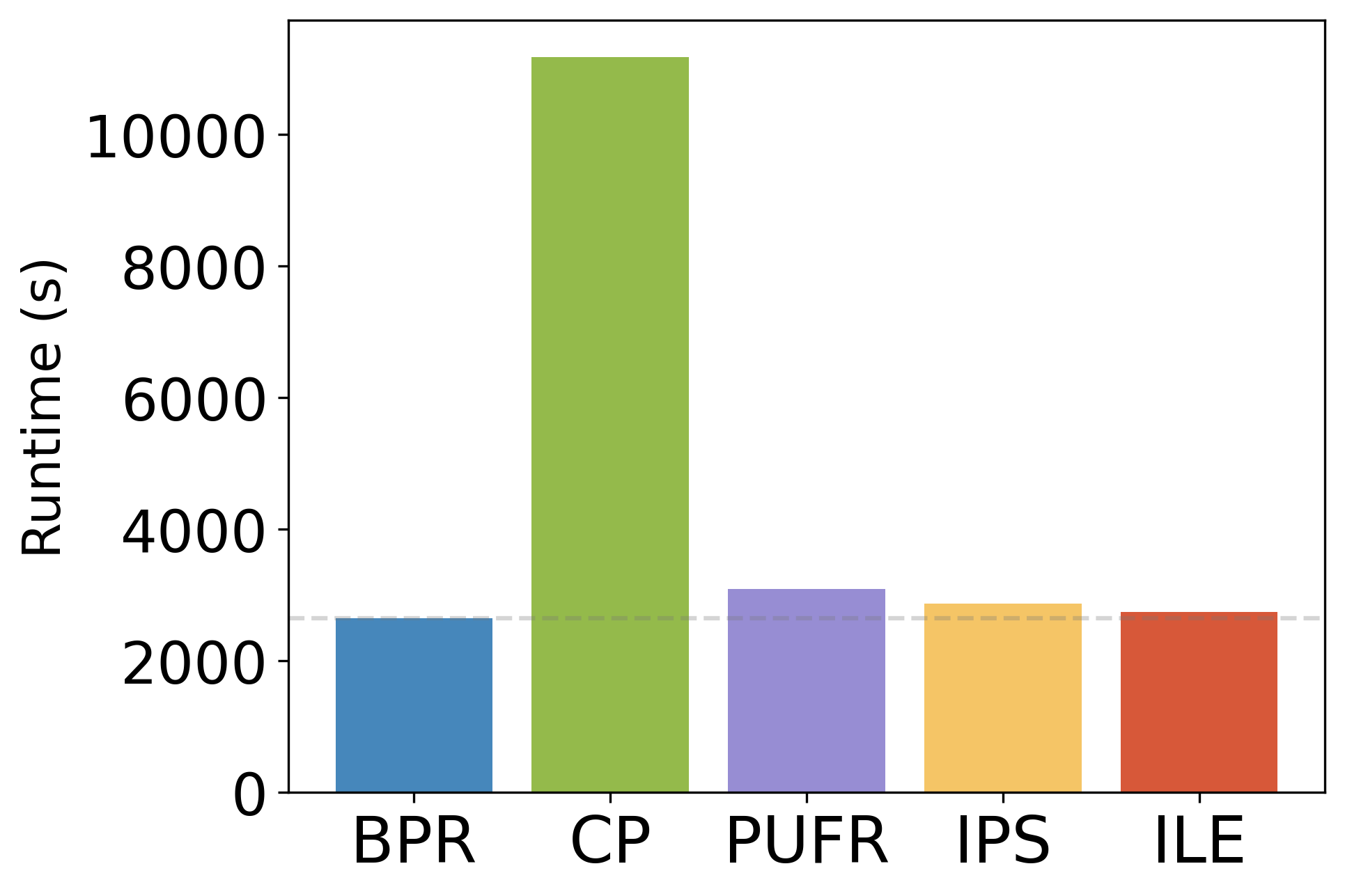}
      \caption{Google Reviews}\label{fig:runtimes googlereview}
    \end{subfigure}\hfill
    \caption{Time complexity of our ILE method and baselines on MovieLens1M, Goodreads, and Google Reviews datasets. For CP and PUFR, the size of the long recommendation lists is 100.}
    \label{fig:runtimes}
\end{figure}

\subsection{Experimental results}

\header{Overall performance}
Table \ref{tab:results} compares the overall performance of our ILE method and baselines on MovieLens1M, Goodreads, and Google Reviews datasets. These results show that our ILE method consistently outperforms all baselines across all metrics and datasets, in particular on Google Reviews dataset that not only it yields higher nDCG, but also outperforms the baselines in terms of fairness metrics. For instance, ILE with $\mathcal{D}=ENT$ on MovieLens1M improves the best baseline (i.e., IPS) by 18.3\%, 8.4\%, and 3.2\% in terms of UPD, AD, and EE, respectively.

\header{Accuracy-Fairness trade-off}
Figure~\ref{fig:fairness-performance} shows performance of our ILE method and baselines in terms of nDCG and fairness metrics for varying values of $\lambda$ on MovieLens1M, Goodreads, and Google Reviews datasets. IPS shows better performance compared to CP and PUFR. However, our ILE method outperforms IPS in general across all fairness metrics and datasets. In particular, for $ILE (\mathcal{D}=ENT)$ on MovieLens1M dataset, for $ILE (\mathcal{D}=STD)$ on Goodreads dataset, and for $ILE (\mathcal{D}=STD)$ and $ILE (\mathcal{D}=MAD)$ on Google Reviews dataset.

\header{Time complexity}
Figure~\ref{fig:runtimes} compares the running time for our ILE method and the baselines. The plot shows that CP has the highest running time. This is due to the extra optimization process in post-processing the long recommendation list. Although PUFR is also a post-processing approach, it does not require heavy processing for generating the recommendation lists, given the uncertainty values are known. IPS and ILE are in-processing methods and their running time depends on the training process and model convergence. While these two methods are competing on the running time, our ILE shows more robust efficiency compared to IPS which confirms its potential for practical application. 

\section{Conclusion}
In this paper, we studied the problem of disparate optimization process across different groups of items. We observed that due to the popularity bias, the convergence rate across different item groups varies. To address this issue, we proposed the \textit{item loss equalization} method that integrates a constraint into the objective function of the recommendation models to minimize the loss disparity among different groups of items during the training process. Experiments on three real-world datasets revealed the effectiveness of our method in mitigating popularity bias compared to the baselines. For future work, we plan to extend our proposed method for addressing other types of bias such as positivity bias and mainstream bias.


\bibliographystyle{splncs04}
\bibliography{ref}

\end{document}